\documentclass[twocolumn,floatfix,floats,aps,prb,showpacs,amsfonts,amssymb]{revtex4}

\usepackage{amssymb}
\usepackage{dcolumn}
\usepackage{bm}

\usepackage{hyphenat}
\usepackage{graphicx}
\usepackage{psfrag}
\usepackage[latin1]{inputenc}
\begin{document}

\title{Interband, intraband and excited-state direct photon absorption of silicon 
and germanium nanocrystals embedded in a wide band-gap lattice}

\author{C. Bulutay}
\email{bulutay@fen.bilkent.edu.tr}
\affiliation{Department of Physics and UNAM - Institute of Materials Science and Nanotechnology, 
Bilkent University, Bilkent, Ankara, 06800, Turkey}
\date{\today}

\begin{abstract}
Embedded Si and Ge nanocrystals (NCs) in wide band-gap matrices are studied 
theoretically using an atomistic pseudopotential approach. 
From small clusters to large NCs containing on the 
order of several thousand atoms are considered. Effective band-gap values as a function 
of NC diameter reproduce very well the available experimental and theoretical data. 
It is observed that the highest occupied molecular orbital for both Si and Ge NCs and 
the lowest unoccupied molecular orbital for Si NCs display oscillations with respect to size 
among the different irreducible representations of the $C_{3v}$ point group to which 
these spherical NCs belong. Based on this electronic structure, first the 
interband absorption is thoroughly studied which shows the importance of surface polarization 
effects that significantly reduce the absorption when included. This reduction is found to 
increase with decreasing NC size or with increasing permittivity 
mismatch between the NC core and the host matrix.
Reasonable agreement is observed with the experimental absorption spectra where available. 
The deformation of spherical NCs into prolate or oblate ellipsoids are seen 
to introduce no pronounced effects for the absorption spectra. Next, intraconduction 
and intravalence band absorption coefficients are obtained in the wavelength range 
from far-infrared to visible region. These results can be valuable for the infrared 
photodetection prospects of these NC arrays. Finally, excited-state absorption at three 
different optical pump wavelengths, 532~nm, 355~nm and 266~nm are studied for 3- and 
4~nm-diameter NCs. This reveals strong absorption windows in the case of holes and a broad 
spectrum in the case of electrons which can especially be relevant 
for the discussions on achieving gain in these structures.
\end{abstract}


%73.22.-f Electronic structure of nanoscale materials: clusters, nanoparticles, nanotubes, and nanocrystals
%78.67.Bf [Optical Properties of] Nanocrystals and nanoparticles
%78.40.-q Absorption and reflection spectra: visible and ultraviolet 

\maketitle

\section{Introduction}
The field of silicon and germanium nanocrystals (NCs) is very active due to 
important technological achievements and prospects particularly in connection 
with optics such as light emitting diodes and lasers.\cite{ossicini,pavesi00,walters,rong} 
Two fundamental processes describing the interaction of light with matter are 
the photon absorption and emission. In the context of NCs, it has been shown 
both experimentally~\cite{wolkin} and theoretically~\cite{zhou,luppi} that 
the interface properties have dramatic effects on the emission 
properties. On the other hand, absorption measurements are less sensitive to surface quality 
and allow for a more direct characterization of the intrinsic structure of 
NCs.\cite{sagnes} Therefore, the study of the direct photon absorption in NCs can  
provide a clear physical understanding. Moreover, with the ever-growing importance of 
renewable energy resources, the research on the new-generation photovoltaics has gained 
momentum and hence the subject of direct photon absorption in nanocrystalline silicon 
($nc$-Si).\cite{pv} 
However, both experimentally~\cite{furukawa,kanemitsu,takeoda,wilcoxon99,kovalev,wilcoxon01} and 
theoretically,~\cite{vasiliev,weissker02a,weissker02b,weissker03,ramos,melnikov,tsolakidis,trani05} 
researchers till now have predominantly focused on the interband absorption process. 
This is the only optical absorption possibility for an intrinsic 
semiconductor NC under equilibrium.
By relaxing these two constraints we can introduce other absorption channels, mainly through 
carrier injection or optical pumping. The associated absorption in either case is sometimes 
referred to as ``free'' carrier absorption despite the carrier confinement in NCs. In our work 
we discriminate between the two. The electrical injection or doping gives rise to intraband 
absorption, also termed as intersubband absorption~\cite{lin} which has 
practical importance for mid- and near-infrared photodetectors.\cite{qdip} The optical pumping 
which is usually well above the effective band gap
leads to excited-state absorption (also termed as photoinduced absorption) which is an undesired 
effect that can inhibit the development of optical gain.\cite{malko}
Recent experiments on excited-state absorption concluded that more attention should be devoted 
to the role of the excitation conditions in the quest for the 
silicon laser.\cite{elliman,trojanek,forcales}
Therefore, the aim of this work is to provide a comprehensive theoretical account of all of these 
direct photon absorption mechanisms in Si and Ge NCs under various size, shape and excitation 
conditions. This provides a complementary track to the existing experimental efforts where 
the size and shape control are currently major obstacles.

The absorption coefficient of the semiconductor NCs depends on 
the product of the optical transition oscillator strength and their joint density of 
states as well as to their volume filling factor within the matrix. Therefore, the essential 
decision on a theoretical study is the sophistication level of the electronic 
structure. The usual trade off between the computational cost and accuracy is operational.
The constraints on the former are quite stringent as a NC including the active region of the 
matrix surrounding itself can contain on the order of ten thousand atoms. 
As for the latter, not only the accuracy but also the validity of a chosen approach can become 
questionable. Computationally low-cost approaches like the envelope function in conjunction with 
8-band {\bf k}$\cdot${\bf p} are not as accurate for this task and furthermore, they miss 
some critical symmetries of the underlying lattice.~\cite{zunger} On the other extreme, there 
lies the density functional theory-based \textit{ab initio} codes~\cite{martin-book} which 
have been applied to smaller NCs containing less than 1000 atoms which still require 
very demanding computational 
resources.~\cite{ogut,weissker02a,weissker02b,weissker03,ramos,luppi,melnikov,tsolakidis} 
The \textit{ab initio} analysis of larger NCs of sizes between 3-10~nm is practically not 
possible with the current computer power.
While this technological hurdle will be gradually overcome in the years to come, 
there exists other atomistic approaches that can be 
employed for NC research which can be run on modest platforms and are much simpler to 
develop, such as the tight binding technique which has been successfully employed 
by several groups.~\cite{delerue04,ren,niquet,trani05} 
On the pseudopotential-based approaches, two new recipes were proposed by Wang and Zunger 
over the last decade.~\cite{wang94a,wang97,wang99} 
The folded spectrum method~\cite{wang94a} relies on standard plane wave basis and direct 
diagonalization; its speed is granted from being focused on relatively few targetted states. 
For the study of excitons this approach becomes very suitable whereas for the optical 
absorption spectra where a large number of states contribute it loses its advantage. 
Their other recipe is the so-called linear combination of bulk bands 
(LCBB).~\cite{wang94a,wang97,wang99} As a matter of fact, the idea of using bulk Bloch 
states in confined systems goes back to earlier times, one of its first implementations being 
the studies of Ninno \emph{et al.}~\cite{ninno85,ninno86} Up to now, it has been used for self-assembled 
quantum dots,~\cite{wang97,wang99} superlattices,~\cite{botti01,botti04} and high-electron mobility 
transistors,~\cite{chirico} and very recently on the $nc$-Si aggregation stages.~\cite{bulutay07} 
In this work, we apply LCBB to the electronic structure and absorption spectra of Si and Ge NCs.
An important feature of this work, in contrast to 
commonly studied hydrogen-passivated NCs is that we consider NCs \textit{embedded} in a wide band-gap matrix which 
is usually silica.~\cite{pecvd} In principle, other matrices such as alumina or silicon nitride can 
be investigated along the same lines.

The organization of the paper is as follows: in Section~II we describe the theoretical framework 
which includes some brief information on the LCBB technique and the absorption expressions. A 
self-critique of the 
theoretical model is done in Section~III. Section~IV presents the results and discussions on the 
band edge electronic structure, interband, intraband, and excited-state absorptions followed by 
our conclusions in Section~V. Appendix section contains technical details on the employed 
pseudopotential form factors and our LCBB implementation.

\section{Theory}
For the electronic structure of large-scale atomistic systems Wang and Zunger 
have developed the LCBB method which is particularly convenient for embedded NCs containing 
several thousand atoms.~\cite{wang97,wang99} 
The fact that it is a pseudopotential-based method makes it more preferable over the 
empirical tight binding technique for the study of optical properties as aimed in this work.
In this technique the NC wavefunction with a state label $j$ is expanded
in terms of the bulk Bloch bands of the constituent core and/or embedding medium (matrix) 
materials
\begin{equation}
\psi_j(\vec{r})=\frac{1}{\sqrt{N}}\sum_{n,\vec{k},\sigma} C^{\sigma}_{n,\vec{k},j}\,
e^{i\vec{k}\cdot\vec{r}} u^{\sigma}_{n,\vec{k}}(\vec{r})  \, ,
\end{equation}
where $N$ is the number of primitive cells within the computational supercell, 
$C^{\sigma}_{n,\vec{k},j}$ is the expansion coefficient set to be determined 
and $\sigma$ is the constituent bulk material label pointing to the NC core or 
embedding medium. $u^{\sigma}_{n,\vec{k}}(\vec{r})$ is the cell-periodic part 
of the Bloch states which can be expanded in terms of the reciprocal lattice 
vectors $\{\vec{G}\}$ as
\begin{equation}
 u^{\sigma}_{n,\vec{k}}(\vec{r})=\frac{1}{\Omega_0}\sum_{\vec{G}}
 B^{\sigma}_{n\vec{k}}\left(\vec{G}\right)e^{i\vec{G}\cdot\vec{r}}\, ,
\end{equation}
where $\Omega_0$ is the volume of the primitive cell.
The atomistic Hamiltonian for the system is given by
\begin{equation}
\label{Hamiltonian}
\hat{H}=-\frac{\hbar^2\nabla^2}{2m}+
\sum_{\sigma,\vec{R}_j,\alpha} W^{\sigma}_{\alpha}(\vec{R}_j)\,
\upsilon^{\sigma}_{\alpha}\left( \vec{r}-\vec{R}_j-\vec{d}^{\sigma}_{\alpha}\right) \, ,
\end{equation}
where $W^{\sigma}_{\alpha}(\vec{R}_j)$ is the weight function that takes values 
0 or 1 depending on the type of atom at the position 
$\vec{R}_j-\vec{d}^{\sigma}_{\alpha}$,~\cite{note1}
and $\upsilon^{\sigma}_{\alpha}$ is the screened spherical pseudopotential 
of atom $\alpha$ of the material $\sigma$. We use semiempirical pseudopotentials for Si and Ge 
developed particularly for strained Si/Ge superlattices which reproduces a large variety of 
measured physical data such as bulk band structures, deformation potentials, electron-phonon 
matrix elements, and heterostructure valence band offsets.~\cite{friedel} With such a choice, 
this approach benefits from the empirical pseudopotential method (EPM), which in addition to its 
simplicity has another advantage over the more accurate density functional \textit{ab initio} techniques 
that run into well-known band-gap problem~\cite{martin-book} 
which is a disadvantage for the correct prediction of the excitation energies.

The formulation can be cast into the following generalized eigenvalue equation:~\cite{wang99,chirico} 

\begin{equation}
\sum_{n,\vec{k},\sigma}H_{n'\vec{k}'\sigma',n\vec{k}\sigma}\,
C^{\sigma}_{n,\vec{k}} =E \sum_{n,\vec{k},\sigma}S_{n'\vec{k}'
\sigma',n\vec{k}\sigma}\,C^{\sigma}_{n,\vec{k}} \, ,
\end{equation}
where 
$$
H_{n'\vec{k}'\sigma',n\vec{k}\sigma}  \equiv  \left\langle n'\vec{k}
'\sigma'\vert\hat{T}+\hat{V}_{\mbox{\begin{scriptsize}xtal\end{scriptsize}}}\vert 
n\vec{k}\sigma \right\rangle\, ,
$$
$$
\left\langle n'\vec{k}
'\sigma'\vert\hat{T}\vert n\vec{k}\sigma \right\rangle  =  \delta_{\vec{k}',\vec{k}}
\sum_{\vec{G}}\frac{\hbar^2}{2m}\left\vert \vec{G}+\vec{k} \right\vert^2 
B^{\sigma'}_{n'\vec{k}'}
\left(\vec{G}\right)^* 
B^{\sigma}_{n\vec{k}}\left(\vec{G}\right)\, , 
$$
\begin{eqnarray}
\left\langle n'\vec{k}
'\sigma'\vert\hat{V}_{\mbox{\begin{scriptsize}xtal\end{scriptsize}}}\vert n\vec{k}\sigma 
\right\rangle & = & \sum_{\vec{G},\vec{G}'}
 B^{\sigma'}_{n'\vec{k}'} \left(\vec{G}\right)^* B^{\sigma}_{n\vec{k}}\left(\vec{G}\right)
\nonumber \\  & &\times  
 \sum_{{\sigma''},\alpha}V_\alpha^{{\sigma''}}\left( \left\vert \vec{G}+\vec{k}-\vec{G}
 '-\vec{k}'\right\vert^2\right) \nonumber \\
  & &\times W_\alpha^{{\sigma''}}\left(\vec{k}-\vec{k}'\right)
 e^{i\left(\vec{G}+\vec{k}-\vec{G}'-\vec{k}'\right)\cdot\vec{d}_\alpha^{{\sigma''}}}
 \, , \nonumber
\end{eqnarray}
$$
S_{n'\vec{k}'\sigma',n\vec{k}\sigma}  \equiv 
\left\langle n'\vec{k}
'\sigma'\vert n\vec{k}\sigma \right\rangle\, .
$$
Here, the atoms are on regular sites of the underlying Bravais lattice: 
$\vec{R}_{n_1,n_2,n_3}=n_1\vec{a}_1+n_2\vec{a}_2+n_3\vec{a}_3$ where $\{\vec{a}_i\}$ are its 
direct lattice vectors of the Bravais lattice. Both the NC and the host matrix are
assumed to possess the same lattice constant and the whole structure is 
within a supercell which imposes the periodicity condition
$W\left(\vec{R}_{n_1,n_2,n_3}+N_i\vec{a}_i\right)=W\left(\vec{R}_{n_1,n_2,n_3}\right)$, 
recalling its Fourier representation $W\left(\vec{R}_{n_1,n_2,n_3}\right)\to\sum 
\tilde{W}(q)e^{i\vec{q}\cdot\vec{R}_{n_1,n_2,n_3}}$, implies $e^{i\vec{q}\cdot N_i\vec{a}_i}=1$, 
so that $\vec{q}\to\vec{q}_{m_1,m_2,m_3}=\vec{b}_1\frac{m_1}{N_1}+\vec{b}_2\frac{m_2}{N_2}
+\vec{b}_3\frac{m_3}{N_3}$, where $\{\vec{b}_i\}$ are the reciprocal lattice vectors of the 
\textit{bulk} material. Thus the reciprocal space of the supercell arrangement is not a 
continuum but is of the grid form composed of points $\{\vec{q}_{m_1,m_2,m_3}\}$, where
$m_i=0,1,\ldots ,N_i-1$.

\begin{figure}[htb]
\includegraphics[width=8cm]{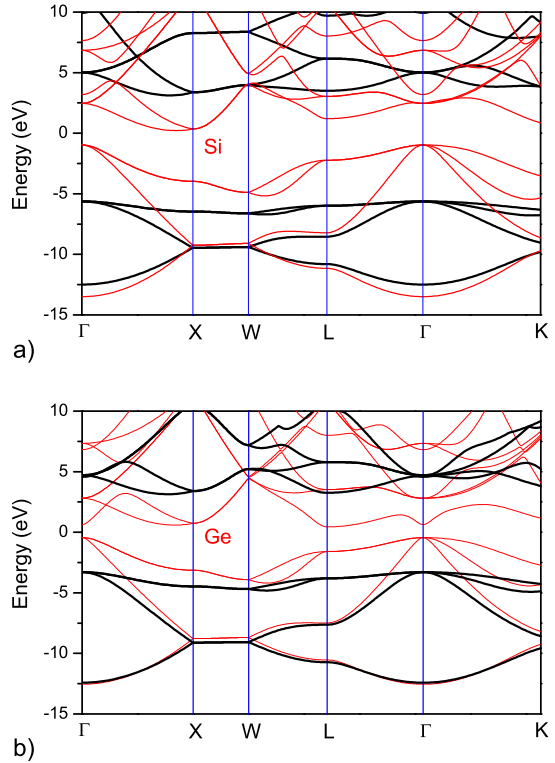}
\caption{\label{fig1}(Color online) EPM band structures for bulk (a) Si, (b) Ge together with their wide 
band-gap matrices (thick lines) which for the former reproduces the band line-up of the Si/SiO$_2$ 
interface.}
\end{figure}

\begin{figure}[htb]
\includegraphics[width=8.5cm]{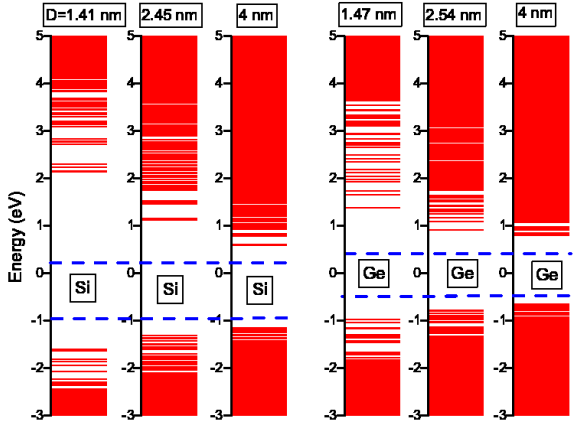}
\caption{\label{fig2}(Color online) The variation of NC states with respect to diameter for Si and Ge NCs.
The bulk band edges are marked with a dashed line for comparison.}
\end{figure}

An important issue is the choice of the host matrix material. If the NC is surrounded by vacuum, 
this corresponds to the free-standing case. However, the dangling bonds of the surface NC atoms 
lead to quite a large number of interface states which adversely contaminate especially 
the effective band-gap region of the NC. In practice NCs are embedded into a wide band-gap host 
matrix which is usually silica.~\cite{pecvd} However, the pseudopotential for oxygen is
nontrivial in the case of EPM~\cite{chelikowsky77} and furthermore, lattice constant of SiO$_2$ is 
not matched to either of the core materials introducing strain effects. 
Therefore, we embed the Si and Ge NCs into an \emph{artificial} wide band-gap 
medium which for the former reproduces the proper band alignment of the Si/SiO$_2$ system. 
To circumvent the strain effects which are indeed present in the actual samples,
we set the lattice constant and crystal structure of the matrix equal to that of the core 
material. 
The pseudopotential form factors of the wide band-gap matrices for Si and Ge can 
easily be produced starting from those of the core materials. More details are provided 
in the Appendix section.
The resultant bulk band structures for Si and Ge and their host wide band-gap matrices are shown 
in Fig.~\ref{fig1}. With the use of such a lattice-matched matrix providing the perfect 
termination of the surface bonds of the NC core atoms lead to the removal of all gap states 
as can be observed in Fig.~\ref{fig2}. In these plots, the evolution of the 
effective band-gaps towards their bulk values (marked by dashed lines) is clearly seen 
as the diameter increases.

\begin{figure}[htb]
\includegraphics[width=8.5cm]{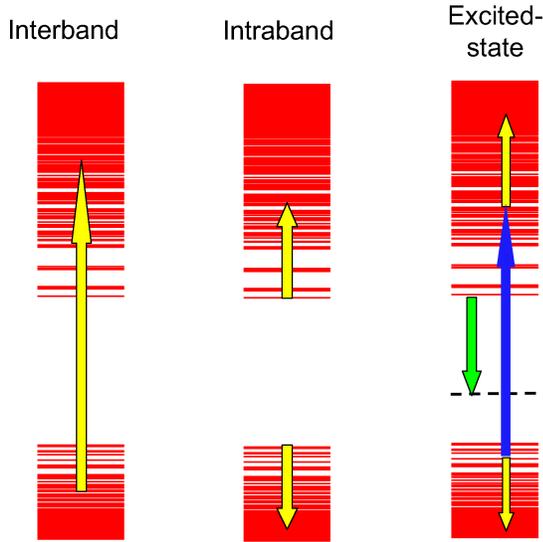}
\caption{\label{fig3}(Color online) Illustration for the three different absorption processes in 
NCs considered in this work: interband, intraband and excited-state absorption. 
The yellow (light-colored) arrows indicate the direct photon absorption transitions, 
the blue (dark-colored) arrow represents optical pumping and the downward green arrow 
corresponds to luminescence which can be to a interface state (dashed line).}
\end{figure}

Once the electronic wavefunctions of the NCs are available, their linear optical properties 
can be readily computed. The three different types of direct (zero-phonon) photon absorption 
processes considered in this work are illustrated in Fig.~\ref{fig3}.
These are interband, intraband and excited-state absorptions. 
In the latter, the blue (dark-colored) arrow represents optical pumping and following carrier 
relaxation, the downward green arrow corresponds to luminescence which can be to a final 
interface state (dashed line).~\cite{pavesi00}
For all these processes, the relevant quantity is the imaginary part of the dielectric function. 
Within the independent-particle 
approximation and the artificial supercell framework~\cite{weissker02a} it becomes 
\begin{eqnarray}
\label{Im_eps}
\mbox{Im}\{\epsilon_{aa}(\omega)\} & = & \frac{\left(2\pi e\hbar\right)^2}{m_0V_{\mbox{\begin{scriptsize}SC\end{scriptsize}}}}
\sum_{c,v}\frac{f^{aa}_{cv}}{E_c-E_v}
\nonumber \\  & & \times 
\frac{\Gamma/(2\pi)}{\left[E_c-E_v-\hbar\omega\right]^2+(\Gamma/2)^2} ,
\end{eqnarray}
where, $a=x,y,z$ denotes the cartesian components of the dielectric tensor and
\begin{equation}
f^{aa}_{cv}=\frac{2m_0\left\vert\langle c\left\vert 
\frac{p_a}{m_0}\right\vert v\rangle \right\vert^2}{E_c-E_v}\, ,
\end{equation}
is the oscillator strength of the transition. In these expressions $m_0$ is the free electron 
mass, $e$ is the magnitude of the electronic charge, and $\Gamma$ is the full-width at half 
maximum value of the Lorentzian broadening. The label $v$ ($c$) correspond to occupied 
(empty) valence (conduction) states 
referring only to their orbital parts in the absence of spin-orbit coupling; the spin summation 
term is already accounted in the prefactor of Eq.~\ref{Im_eps}. 
Finally, $V_{\mbox{\begin{scriptsize}SC\end{scriptsize}}}$ is the volume of the supercell which is a fixed value chosen conveniently large 
to accommodate the NCs of varying diameters, however, if one uses instead, that of the NC, 
$V_{\mbox{\begin{scriptsize}NC\end{scriptsize}}}$, this corresponds calculating $\mbox{Im}\{\epsilon_{aa}\}/f_v$ where 
$f_v=V_{\mbox{\begin{scriptsize}NC\end{scriptsize}}}/V_{\mbox{\begin{scriptsize}SC\end{scriptsize}}}$ 
is the volume filling ratio of the NC. For the sake of generality, this is the form we 
shall be presenting our results.
The electromagnetic \emph{intensity} absorption coefficient $\alpha(\omega)$ is related to the imaginary part of the 
dielectric function through~\cite{jackson} 
\begin{equation}
\label{epsalpha}
\mbox{Im}\{\epsilon_{aa}(\omega)\}=\frac{n_r c}{\omega}\alpha_{aa}(\omega)\, ,
\end{equation}
where $n_r$ is the index of refraction and $c$ is the speed of light.

In the case of intraband absorption, its rate depends on the amount of excited carriers. 
Therefore, we consider the absorption rate for \emph{one} excited electron or hole that lies at an 
initial state $i$ with energy $E_i$. As there are a number of closely spaced such states, 
we perform a Boltzmann averaging over these states as 
$e^{-\beta E_i}/\sum_j e^{-\beta E_j}$. 
We further assume that the final states have no occupancy restriction, 
which can easily 
be relaxed if needed. The expression for absorption rate per an excited 
carrier in each NC becomes
\begin{eqnarray}
\frac{\alpha_{aa}}{f_v} & = & \frac{\pi e^2}{2m_0 c n_r\omega V_{\mbox{\begin{scriptsize}NC\end{scriptsize}}}}
\sum_{i,f}\frac{e^{-\beta E_i}}{\sum_j e^{-\beta E_j}} f^{aa}_{fi}\left[E_f-E_i\right]
\nonumber \\  & & \times 
\frac{\Gamma/(2\pi)}{\left[E_f-E_i-\hbar\omega\right]^2+(\Gamma/2)^2}\, ,
\end{eqnarray}
where again $a$ is the light polarization direction.

Finally, we include the surface polarization effects, also called local field effects (LFE) using 
a simple semiclassical model which agrees remarkably well with more rigorous treatments.~\cite{trani07}
We give a brief description of its  implementation. First, using the expression
\begin{equation}
\label{mixing}
\epsilon_{\mbox{\begin{scriptsize}SC\end{scriptsize}}}=f_v\epsilon_{\mbox{\begin{scriptsize}NC\end{scriptsize}}}+(1-f_v)\epsilon_{\mbox{\begin{scriptsize}matrix\end{scriptsize}}}\, ,
\end{equation}
we extract (i.e., de-embed) the size-dependent NC dielectric function, $\epsilon_{\mbox{\begin{scriptsize}NC\end{scriptsize}}}$, 
where $\epsilon_{\mbox{\begin{scriptsize}SC\end{scriptsize}}}$ corresponds to Eq.~\ref{Im_eps}, suppressing the cartesian indices.
$\epsilon_{\mbox{\begin{scriptsize}matrix\end{scriptsize}}}$ is the dielectric function of the host matrix; for simplicity we set 
it to the permittivity value of SiO$_2$, i.e., $\epsilon_{\mbox{\mbox{\begin{scriptsize}matrix\end{scriptsize}}}}=4$. Since the wide band-gap 
matrix introduces no absorption up to an energy of about 9~eV, we can approximate 
$\mbox{Im}\{\epsilon_{\mbox{\begin{scriptsize}NC\end{scriptsize}}}\}=\mbox{Im}\{\epsilon_{\mbox{\begin{scriptsize}SC\end{scriptsize}}}\}/f_v$. 
One can similarly obtain the $\mbox{Re}\{\epsilon_{\mbox{\begin{scriptsize}NC\end{scriptsize}}}\}$ within the random-phase 
approximation,~\cite{trani05} hence get the
full complex dielectric function $\epsilon_{\mbox{\begin{scriptsize}NC\end{scriptsize}}}$. According to the classical Clausius-Mossotti 
approach, which is shown to work also for NCs,~\cite{mahan} the dielectric function of the NC is modified 
as
\begin{equation}
\label{LFE}
\epsilon_{\mbox{\begin{scriptsize}NC,LFE\end{scriptsize}}}=\epsilon_{\mbox{\begin{scriptsize}matrix\end{scriptsize}}}\left[ \frac{4\epsilon_{\mbox{\begin{scriptsize}NC\end{scriptsize}}}-
\epsilon_{\mbox{\begin{scriptsize}matrix\end{scriptsize}}}}{\epsilon_{\mbox{\begin{scriptsize}NC\end{scriptsize}}}+2\epsilon_{\mbox{\begin{scriptsize}matrix\end{scriptsize}}}}  \right]\, ,
\end{equation}
to account for LFE. The corresponding supercell dielectric function, $\epsilon_{\mbox{\begin{scriptsize}SC,LFE\end{scriptsize}}}$ follows 
using Eq.~\ref{mixing}. Similarly, the intensity absorption coefficients are also modified due to surface 
polarization effects, cf.~Eq.~\ref{epsalpha}. Its consequences will be reported in Section~IV.

\section{A self-critique of the theoretical model}
The most crucial simplification of our model is the fact that strain-related effects are 
avoided, a route which is shared by other theoretical 
works.~\cite{delley,weissker02a,weissker02b,niquet,trani05} 
For large NCs this may not be critical, however, for very small sizes this simplification is 
questionable. An important support for our act is that Weissker and coworkers have concluded 
that while there is some shift and possibly a redistribution of oscillator strengths after 
ionic relaxation, the overall appearance of the absorption spectra does not change 
strongly.~\cite{weissker03} 
We should mention that Wang and Zunger have offered a recipe for including strain within the 
LCBB framework, however, this is considerably more involved.~\cite{wang99}
Another widespread simplification on Si and Ge NCs is the omission 
of the spin-orbit coupling and the nonlocal (angular momentum-dependent) pseudopotential terms in 
the electronic structure Hamiltonian. Especially the former is not significant for Si which 
is a light atom but it can have a quantitative impact on the valence states of Ge NCs; 
such a treatment is available in Ref.~\onlinecite{reboredo01}.

On the dielectric response, there are much more sophisticated and involved treatments~\cite{onida} 
whereas ours is equivalent to the independent particle random phase 
approximation~\cite{ehrenreich} of the macroscopic dielectric function with the surface polarization 
effects included within the classical Clausius-Mossotti model.~\cite{trani07} The contribution 
of the excluded excitonic and other many-body effects beyond the mean-field level can be assessed 
\emph{a posteriori} by comparing 
with the available experimental data. However, it is certain that the precedence should be given 
to classical electrostatics for properly describing the background dielectric mismatch between 
the core and the wide band-gap matrix.~\cite{trani07} 
In our treatment this is implemented at an atomistic level.

Another effect not accounted in this work is the role of the interface region. Our wide band-gap 
matrix can reproduce the proper band alignment and dielectric confinement of an SiO$_2$ matrix, 
however, the interface chemistry such as  silicon-oxygen double bonds~\cite{luppi} are not represented. 
These were shown to be much more effective on the emission spectra.~\cite{zhou,luppi} Nevertheless, 
our results can be taken as the benchmark for the performance of the atomistic quantum and 
dielectric confinement with a clean and inert interface. Finally, we do not consider 
the phonon-assisted~\cite{hybertsen} 
or nonlinear absorption. The list of these 
major simplifications also suggest possible improvements of this work.

\section{Results and discussions}
In this section we present our theoretical investigation of the linear optical properties 
of Si and Ge NCs. Three different direct photon absorption processes
are considered as illustrated in Fig.~\ref{fig3} each of which can serve for technological 
applications as well as to our basic understanding.
However, we first begin with the dependence of the optical gap 
on the NC size, mainly as a check of our general framework. 
There exist two different atomic arrangements of a spherical NC depending on whether the center 
of the NC is an atomic position or a tetrahedral interstitial location; under no ionic relaxation, 
Delley and Steigmeier have treated both of these classes as having the $T_d$ point symmetry.~\cite{delley} 
However, the tetrahedral interstitial-centered arrangement should rather have the lower point symmetry of $C_{3v}$ 
and it is the arrangement that we construct our NCs. This leads to even number of NC core atoms, whereas 
it becomes an odd number with the $T_d$ point symmetry. We identify the irreducible representation of a chosen 
NC state by checking its projection to the subspace of each representation.~\cite{reboredo00} For the 
$C_{3v}$ point group these are denoted by $A_1$, $A_2$, and $E$. We utilize this group-theoretic 
analysis in the next subsections.

\begin{figure}[htb]
\includegraphics[width=8cm]{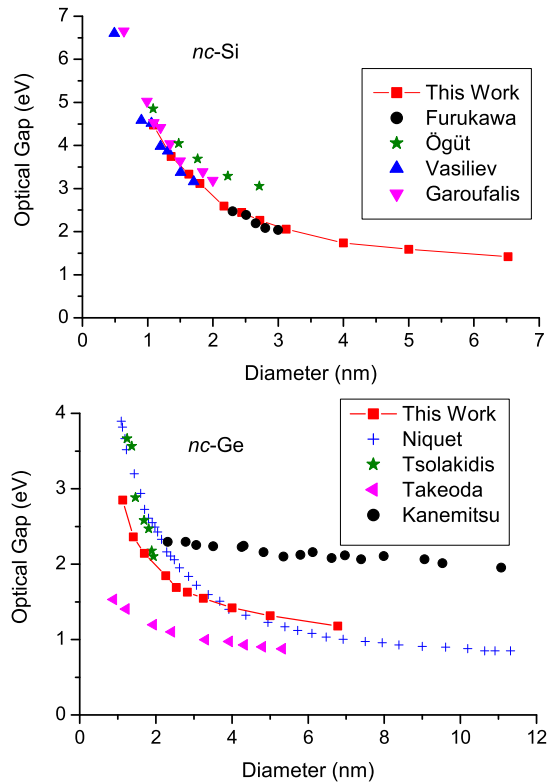}
\caption{\label{fig4}(Color online) Comparison of optical gap as a function of NC diameter of this work 
with previous experimental and theoretical data: Furukawa,~\cite{furukawa} Kanemitsu,~\cite{kanemitsu}
Takeoda,~\cite{takeoda} \"O\u{g}\"ut,~\cite{ogut} Vasiliev,~\cite{vasiliev} Garoufalis,~\cite{garoufalis} 
Niquet,~\cite{niquet} Tsolakidis.~\cite{tsolakidis}}
\end{figure}

\subsection{Effective optical gap}
The hallmark of quantum size effect in NCs has been the effective optical gap with quite 
a number of theoretical~\cite{wang94b,ogut,vasiliev,garoufalis,tsolakidis,trani05} and 
experimental~\cite{kanemitsu,takeoda,wilcoxon99,wilcoxon01} studies performed 
within the last decade. 
Figure~\ref{fig4} contains a compilation of some representative 
results. For Si NCs, it can be observed that there is a good agreement among the existing 
data, including ours. On the other hand, for the case of Ge NCs there is a large spread 
between the experimental data whereas our theoretical results are in very good agreement 
with both \emph{ab initio}~\cite{tsolakidis} and tight binding results.~\cite{trani05}
In our approach the optical gap directly corresponds to the LUMO-HOMO energy difference, as calculated 
by the single-particle Hamiltonian in Eq.~(\ref{Hamiltonian}). This simplicity 
relies on the finding of Delerue and 
coworkers that the self-energy and Coulomb corrections almost exactly cancel each 
other for Si NCs larger than a diameter of 1.2~nm.~\cite{delerue00} 

\begin{figure}[htb]
\includegraphics[width=6cm]{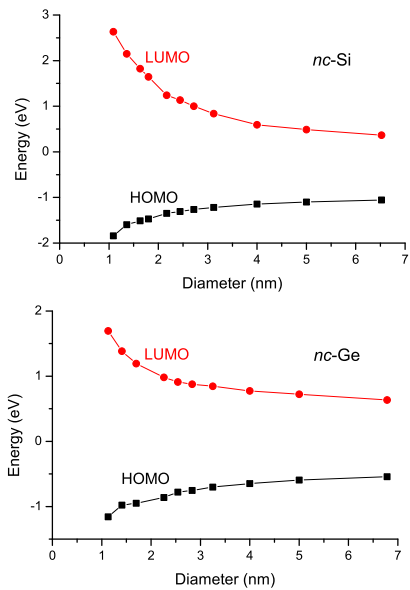}
\caption{\label{fig5}(Color online) The variation of HOMO and LUMO energies with respect to NC diameter 
for Si and Ge NCs that belong to $C_{3v}$ point group.}
\end{figure}
\begin{figure}[htb]
\includegraphics[width=8cm]{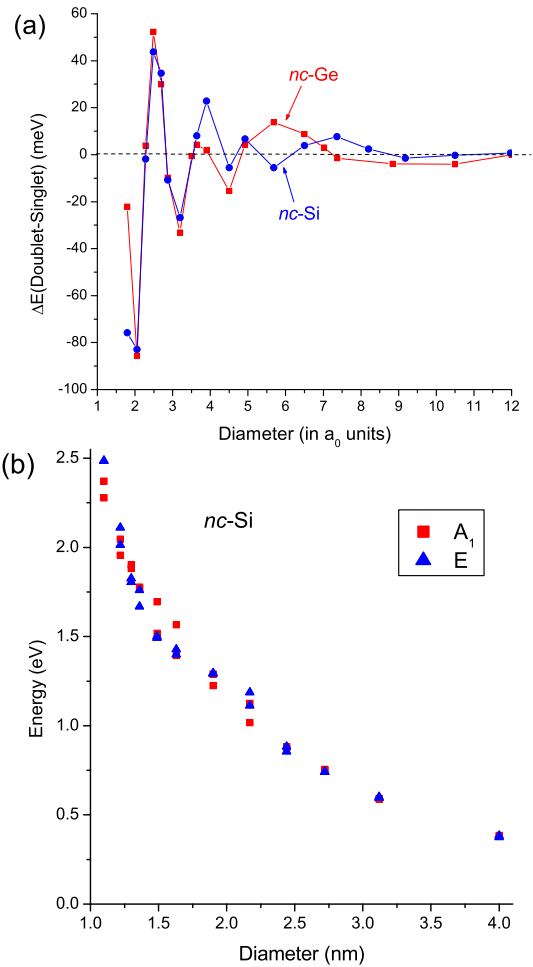}
\caption{\label{fig6}(Color online) (a) The energy difference between doubly degenerate and nondegenerate 
states, one of which becomes the HOMO with respect to diameter in Si and Ge NCs that belong 
to $C_{3v}$ point group; solid lines are for guiding the eyes; $a_0$ is the lattice constant
 for Si or Ge NC. (b)~The lowest three conduction 
states, one of which becomes the LUMO with respect to diameter in Si NCs.}
\end{figure}

\subsection{HOMO and LUMO oscillations with respect to size}
When we plot the variation of individual LUMO and HOMO levels as in Fig.~\ref{fig5} 
we observe with the exception of $nc$-Ge LUMO curve some non-smooth behavior that gets pronounced 
towards smaller sizes. The triple degeneracy in the absence of spin-orbit coupling 
of the valence band maximum in bulk Si and Ge 
is lifted into two degenerate and one nondegenerate 
states. The energy difference between these two set of states is observed to display an 
oscillatory behavior as the NC size gets smaller as shown in Fig.~\ref{fig6}~(a). 
Using the $C_{3v}$ point group symmetry 
operations we identify the doubly degenerate states to belong to $E$ representation and 
nondegenerate state to $A_1$ or $A_2$. Furthermore, we observe a similar oscillation in the 
LUMO region of Si NCs as shown in Fig.~\ref{fig6}(b). The low-lying conduction 
states of Si NCs form six-pack groups which is inhereted from the six equivalent 0.85X 
conduction band minima of \emph{bulk} Si. 
The confinement marginally lifts the degeneracy by sampling contributions from other 
parts of the Brillouin zone. This trend is observed in Fig.~\ref{fig6}(b) as the NC size 
gets smaller.
On the other hand, for Ge NCs all LUMO states 
belong to the same $A_1$ representation and therefore shows no oscillations 
(cf. Fig.~\ref{fig5}). Ultimately, the source of these oscillations is the variation 
of the asphericity of the NCs of $C_{3v}$ point symmetry with respect to size, which 
can energetically favor one of the closely spaced states. In the case of the LUMO 
state of Ge NCs, there is a substantial energy gap between LUMO and the next 
higher-lying state.

\renewcommand{\baselinestretch}{1.5}
\begin{table*}
%Table-HOMO-LUMO-C3v Symmetries
\caption{$C_{3v}$ irreducible representations of the HOMO and LUMO in Si and Ge NCs of various diameters ($D$).}
\begin{ruledtabular} 
\begin{tabular}{c c c c c c c}
			      &    		  & \emph{nc}-Si &		&			& \emph{nc}-Ge &	    \\
\cline{2-4} \cline{5-7}
$N_{\mbox{\small{core}}}$ & $D$ (nm)  &  HOMO	 &	LUMO	&	$D$ (nm)&	HOMO	&	LUMO     \\
\hline                                                      
  32	          &	1.06  	  &	$A_1$	&	$A_1$	&	1.11    &	$A_1$	&	$A_1$    \\
  38	          &	1.13	  &	$A_1$	&	$A_1$	&	1.18    & 	$A_1$	&	$A_1$    \\
  56	          &	1.29	  &	$A_1$	&	$A_1$	&	1.34    & 	$A_1$	&	$A_1$    \\
  74	          &	1.41	  &	$E	$	&	$E	$	&	1.47    &	$E	$	&	$A_1$    \\
  86	          &	1.49	  &	$E	$	&	$E	$	&	1.55    &	$E	$	&	$A_1$    \\
  116	          &	1.64	  &	$A_1$	&	$E	$	&	1.71    &	$A_1$	&	$A_1$    \\
  130	          &	1.71	  &	$E 	$	&   $A_1$	&   1.78    &   $A_2$	&	$A_1$    \\
  136	          &	1.73	  &	$A_1$	&	$E	$	& 	1.80    &	$A_1$	&	$A_1$    \\
  190	          &	1.94	  &	$A_1$	&	$A_1$	&	2.01    &	$A_1$	&	$A_1$    \\
  264	          &	2.16	  &	$E	$	&	$A_1$	&	2.25    &	$E	$	&	$A_1$    \\
  384	          &	2.45	  &	$A_1$	&	$E	$	&	2.55    &	$A_2$	&	$A_1$    \\
  522	          &	2.71	  &	$E	$	&	$E	$	&	2.82    &	$E	$	&	$A_1$    \\
  690	          &	2.98	  &	$A_1$	&	$A_1$	&	3.10    &	$A_2$	&	$A_1$    \\
  768	          &	3.08	  &	$A_1$	&	$A_1$	&	3.21    &	$E	$	&	$A_1$    \\
  1702	          &	4.02	  &	$E	$	&	$E	$	&	4.18    &	$A_2$	&	$A_1$    \\
\end{tabular}
\end{ruledtabular} 
\end{table*} 
\renewcommand{\baselinestretch}{1}

\begin{figure}[htb]
\includegraphics[width=8cm]{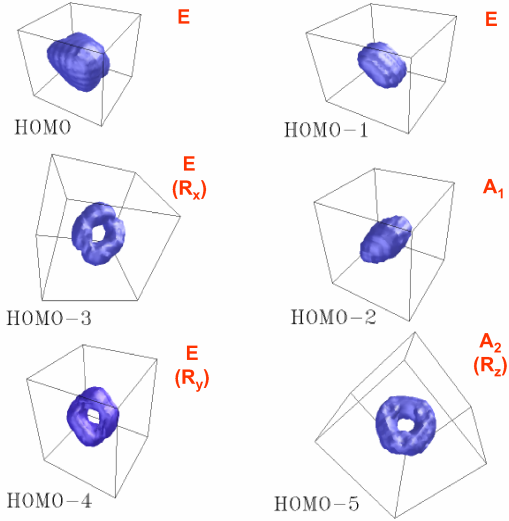}
\caption{\label{fig7}(Color online) The isosurfaces of the envelopes of the wavefunctions of the highest-lying 
six valence states up to HOMO for a Si NC of diameter 2.16~nm. The isosurfaces are drawn 
for the 95\% of the peak value of the envelope wavefunctions.
The $C_{3v}$ point group representations are indicated for each wavefunction. 
Note that some of the plots are rotated with respect to others for best viewing angle.}
\end{figure}

For further insight, we display in Fig.~\ref{fig7} the isosurface 
plots of the envelope of the six highest states 
up to HOMO for a Si NC of diameter 2.16~nm. Point group representation of each 
state is also indicated. For this particular diameter, HOMO has $E$ representation 
which is twofold degenerate. The nondegenerate $A_1$ state also becomes the HOMO 
for different diameters. This is illustrated in Table~I which shows the evolution 
of the HOMO and LUMO symmetries as a function of diameter for Si and Ge NCs. There, 
it can be observed that for the latter the HOMO can also acquire the $A_2$ for larger diameters.

\subsection{Interband absorption}
\begin{figure}[htb]
\includegraphics[width=8cm]{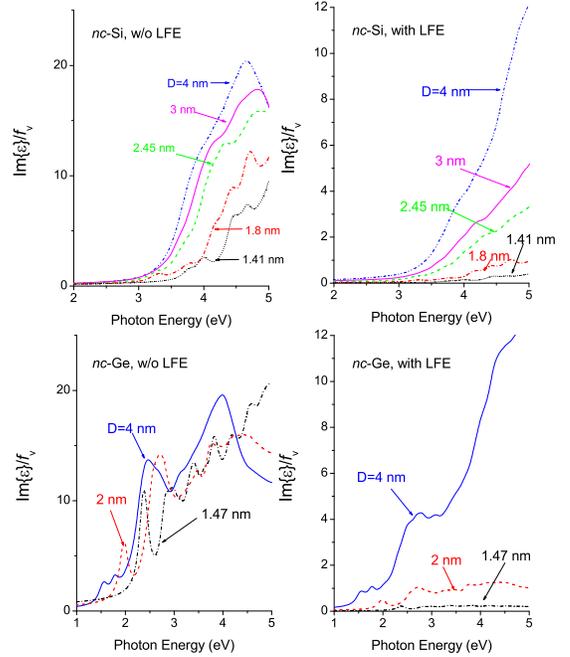}
\caption{\label{fig8}(Color online) The imaginary part of dielectric function for unity volume 
filling factor, $f_v$ for Si and Ge NCs at different diameters with (right panel) and 
without (left panel) local field effects. A Lorentzian broadening energy full width of 200~meV is used.}
\end{figure}

\begin{figure}[htb]
\includegraphics[width=8cm]{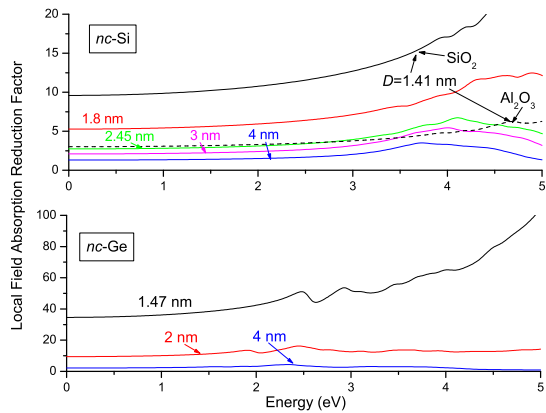}
\caption{\label{fig9}(Color online) The local field absorption factor extracted from the previous figure.}
\end{figure}

The interband absorptions of Si and Ge NCs for a variety of diameters are shown in 
Fig.~\ref{fig8}. For a fair comparison, all different size NCs should possess the 
same volume filling factor. Therefore, we display the results at unity volume 
filling or equivalently per $f_v$.\cite{weissker02b} 
The left and right panels display the cases without and with surface polarization effects 
(or LFE), respectively. There exists remarkable differences between 
the two for both Si and Ge NCs. For instance, even though Ge NCs do not show significant 
size dependence without LFE, this is not the case when LFE is included.
From the ratio of both panels, the so-called local field absorption reduction factor can be 
extracted as shown in Fig.~\ref{fig9}. It can be observed that its size dependence is much 
stronger than the energy dependence.
This reduction in the absorption due to LFE can become a major concern for solar cell applications.
It needs to be mentioned that this effect is highly sensitive to the permittivity mismatch between the core 
and matrix media.
To illustrate this point, in Fig.~\ref{fig9} the case for Al$_2$O$_3$ matrix (having a 
permittivity of 9.1) is also displayed for 1.41~nm Si NC, where it can be seen that compared to 
SiO$_2$ (with a permittivity of about 4) the reduction in absorption due to LFE is much less.
Based on this finding, we employ these size-dependent absorption reduction factors in the 
results to follow including the intraband and excited-state cases.

\begin{figure}[htb]
\includegraphics[width=8cm]{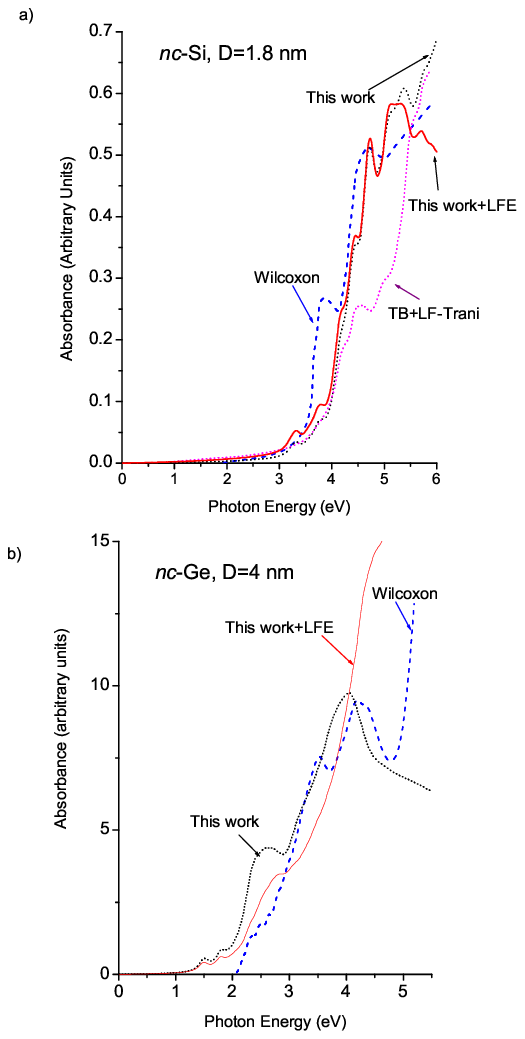}
\caption{\label{fig10}(Color online) Comparison of our absorbance results with the available data: for Si, 
experimental work of Wilcoxon~\cite{wilcoxon99} and the theoretical tight binding 
results of Trani~\cite{trani07} and for Ge, the experimental work of Wilcoxon~\cite{wilcoxon01}.
For our spectra a Lorentzian broadening energy full width of 200~meV is used.}
\end{figure}

In Fig.~\ref{fig10} we compare our results with the experimental data of Wilcoxon 
\emph{et al.} for Si NCs~\cite{wilcoxon99} and Ge NCs.~\cite{wilcoxon01} There is a 
good overall agreement in both cases especially with LFE, however, for the case of Si NCs 
this is much more satisfactory. The major discrepancies can be attributed to excitonic 
effects that are not included in our work.
In the case of Si NCs (Fig.~\ref{fig10}(a)), we also display the tight binding result 
of Trani \emph{et al.} which also includes LFE.~\cite{trani07}

\begin{figure}[htb]
\includegraphics[width=8cm]{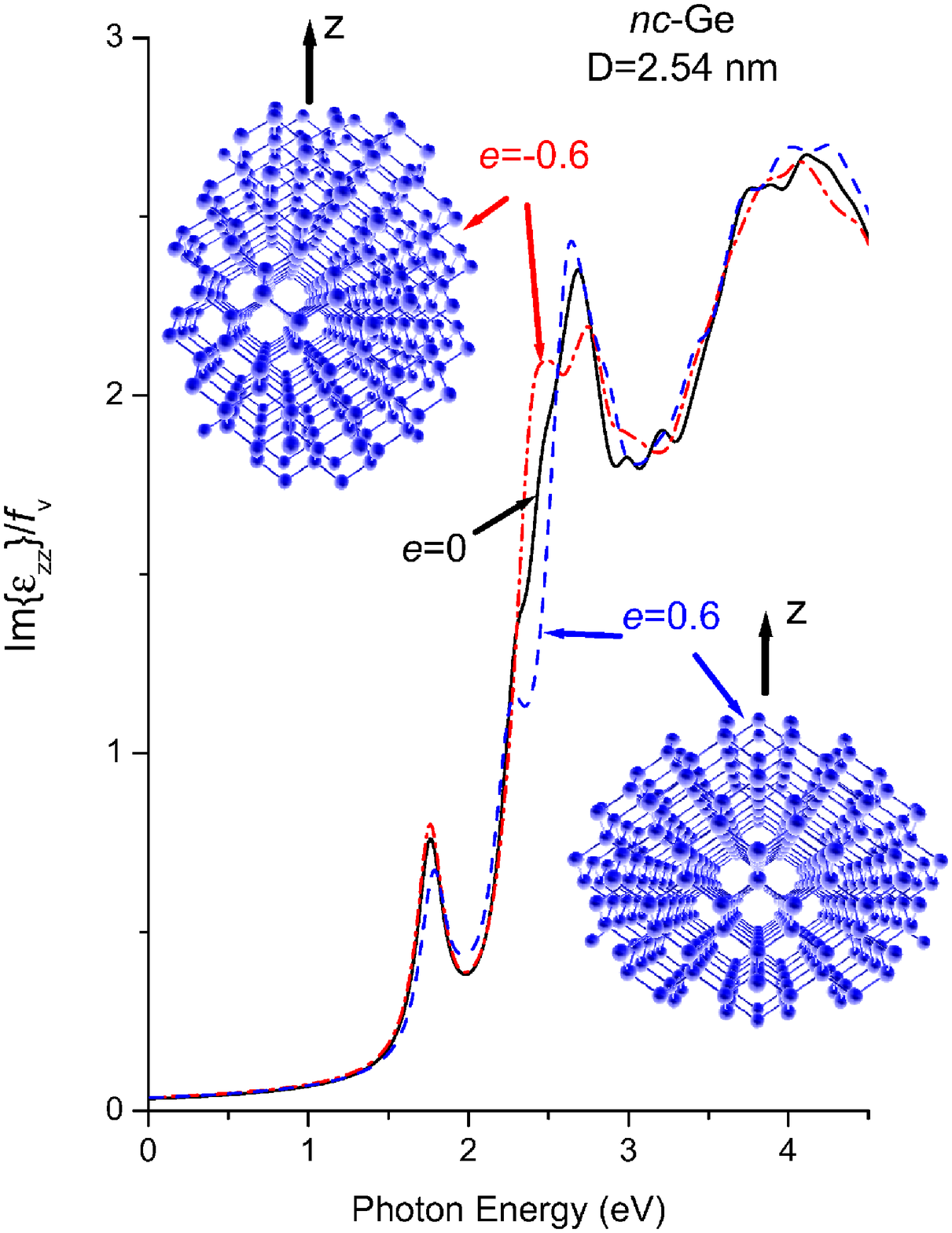}
\caption{\label{fig11}(Color online) The effect of ellipticity, $e$ on the 
$\mbox{Im}\{\epsilon_{zz}\}/f_v$ for a Ge NC with a diameter of 2.54~nm. 
The insets show NC core atoms of the prolate 
($e=-0.6$) and oblate ($e=0.6$) shapes; the $z$-direction is also indicated. 
A Lorentzian broadening energy full width of 200~meV is used.}
\end{figure}

An issue of practical concern is the effect of deviation from the spherical shape 
of the NCs depending on the growth conditions.
At this point we would like to investigate the effect of shape anisotropy on 
the interband absorption. Starting from a spherical 2.54~nm diameter Ge NC, we form 
prolate and oblate ellipsoidal NCs with ellipticities $e=-0.6$ and +0.6, respectively.
All three NCs contain the same number of 384 core atoms; the atomic arrangement of 
the ellipsoidal NCs are shown in the inset of Fig.~\ref{fig11}. In the same figure we 
compare the $zz$ components of the imaginary part of the dielectric tensor for 
three different ellipticities. It is observed that the effect on the interband 
absorption is not significant; the difference is even less for the Si NCs (not shown).

\subsection{Intraband absorption}
\begin{figure}[htb]
\includegraphics[width=9cm]{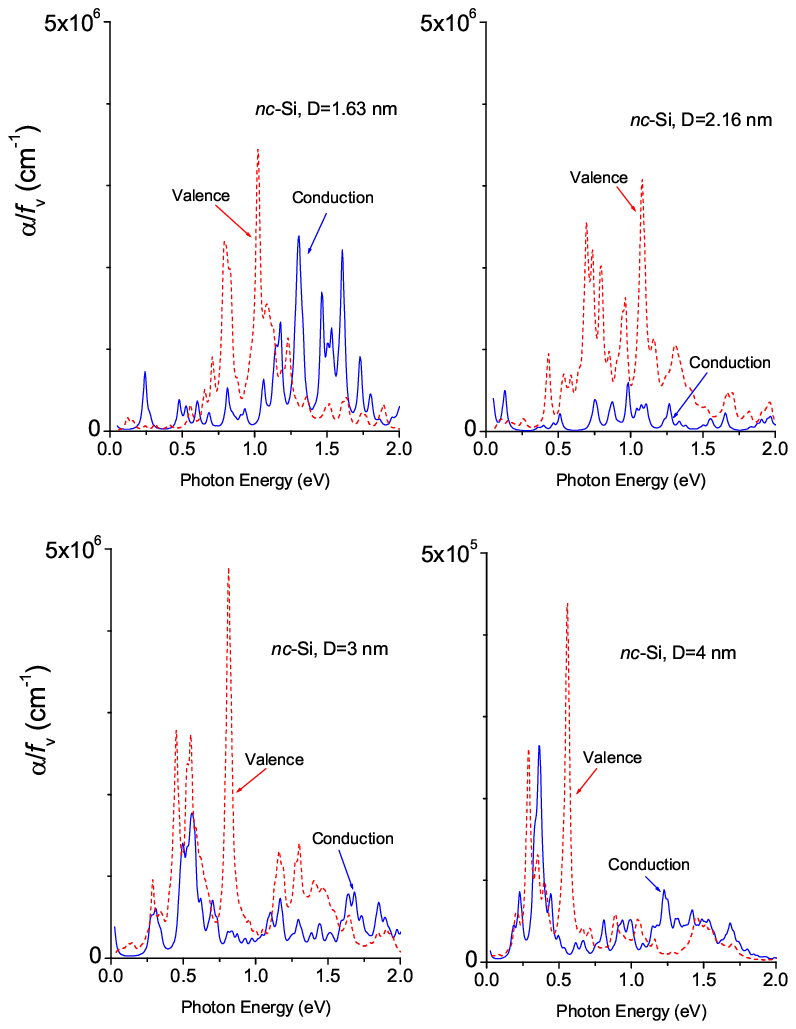}
\caption{\label{fig12}(Color online) Intravalence and intraconduction state absorption coefficients 
in Si NCs of different diameters per excited carrier and at unity filling factor. 
A Lorentzian broadening energy full width of 30~meV is used.
Mind the change in the vertical scale for 4~nm diameter case.}
\end{figure}

\begin{figure}[htb]
\includegraphics[width=9cm]{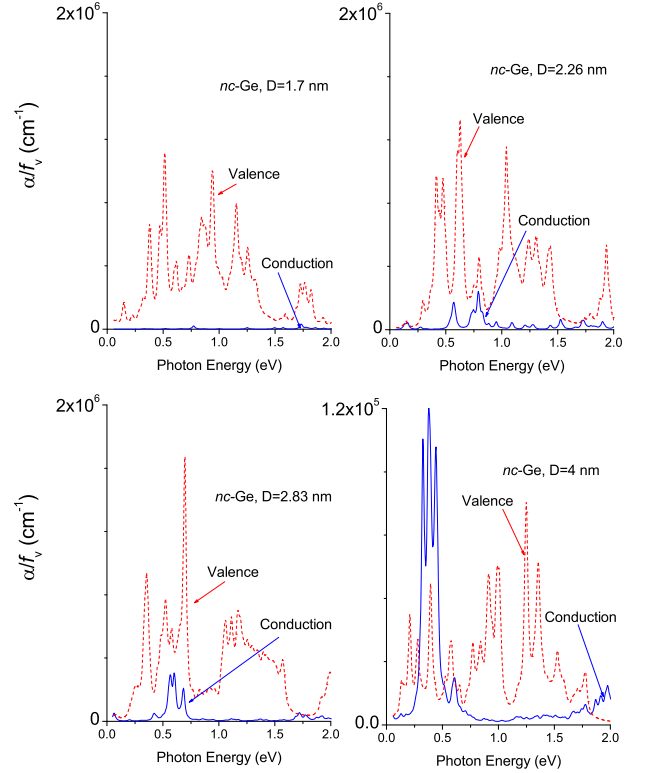}
\caption{\label{fig13}(Color online)  Same as Fig.~\ref{fig12} but for Ge NCs.}
\end{figure}

Unlike the interband case, for the intraband absorption we need to introduce electrons 
to the conduction states or holes to the valence states by an injection mechanism. 
We assume that after injection these carriers relax to their respective band edges 
and attain a thermal distribution. 
Therefore, we perform a Boltzmann averaging at room temperature (300~K) over the 
initial states around LUMO (HOMO) for electrons (holes). The absorption coefficients 
to be presented are for unity volume filling factors and for one carrier per NC; they 
can easily be scaled to a different average number of injected carriers and 
volume filling factors. In Fig.~\ref{fig12} the Si NCs of different diameters are 
compared. The intraband absorption is observed to be enhanced as the NC size grows up to 
about 3~nm followed by a drastic fall for larger sizes. For both holes and electrons very 
large number of absorption peaks are observed from 0.5~eV to 2~eV.
Recently, de Sousa \emph{et al.} have also considered the intraband absorption in Si 
NCs using the effective mass approximation and taking into account the multi-valley anisotropic 
band structure of Si.\cite{desousa} However, their absorption spectra lacks much of the 
features seen in Fig.~\ref{fig12}.
Turning to Ge NCs, shown in Fig.~\ref{fig13} the intravalence band absorption 
profile is very similar to that of Si NCs, however in this case the intraconduction band 
absorption is much weaker.

Mimura \emph{et al.} have measured the optical absorption in heavily 
phosphorus doped Si NCs of a diameter of 4.7~nm.~\cite{mimura} This provides us an 
opportunity to compare our results on the intraconduction band absorption in Si NCs. 
There is a good order-of-magnitude agreement. However, in contrast to our spectra 
in Fig.~\ref{fig12} which contains well-resolved peaks, they have registered a smooth 
spectrum which has been attributed by the authors to the smearing out due to size 
and shape distribution within their NC ensemble.~\cite{mimura}

\subsection{Excited-state absorption}
Finally, we consider another intraband absorption process where the system is under a 
continuous interband optical pumping that creates electrons and holes with excess energy. 
We consider three different excitation wavelengths: 532~nm, 355~nm and 266~nm 
which respectively correspond to the second-, third- and fourth-harmonic of 
the Nd-YAG laser at 1064~nm. The initial states of 
the carriers after optical pumping are chosen to be at the pair of states with the
maximum oscillator strength~\cite{bulutay07} among interband transitions under the chosen excitation.
The determined energies of these states are tabulated in Table~II where it can be 
observed that in general the excess energy is unevenly partitioned, mainly in favor of 
the conduction states.  
Once again a Boltzmann averaging is used to get the 
contribution of states within the thermal energy neighborhood. 

\begin{figure}[htb]
\includegraphics[width=8.5cm]{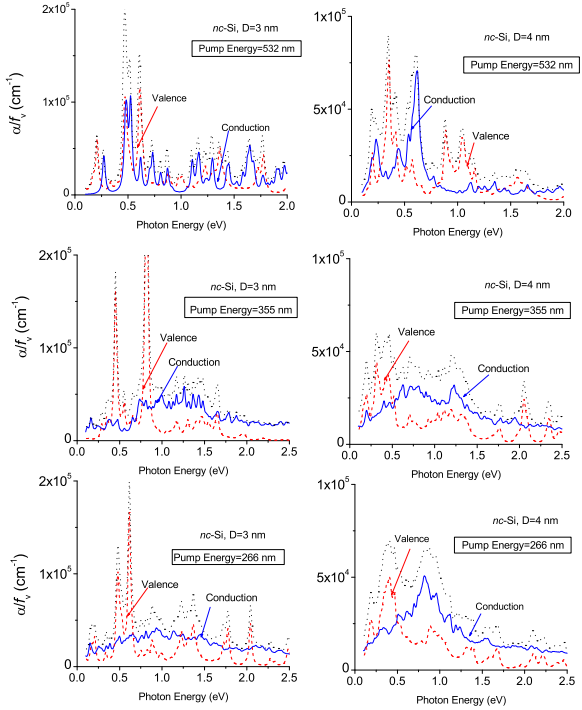}
\caption{\label{fig14}(Color online) Excited-state absorption within valence and conduction 
states of Si NCs per excited carrier and at unity filling factor under 
three different optical pumping wavelengths of 532~nm, 
355~nm and 266~nm. Dotted lines in black color refer to total absorption coefficients. 
Two different diameters are considered, 3 and 4~nm.
A Lorentzian broadening energy full width of 30~meV is used.}
\end{figure}

\begin{figure}[htb]
\includegraphics[width=8.5cm]{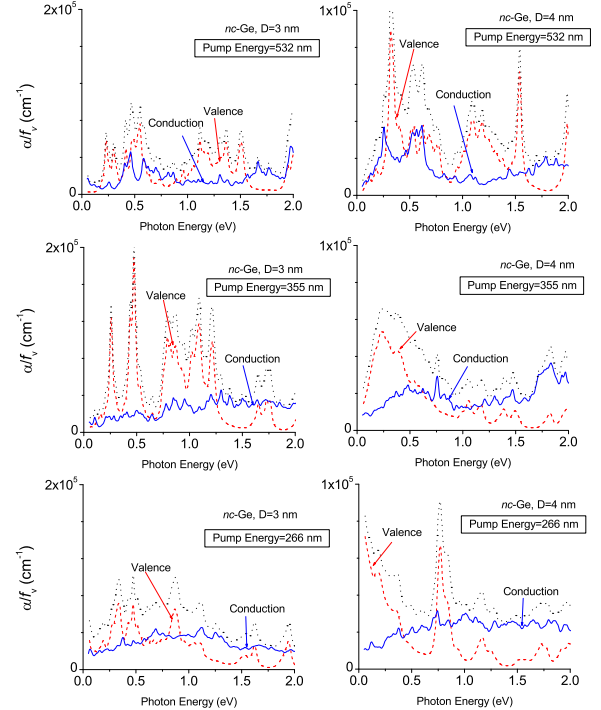}
\caption{\label{fig15}(Color online) Same as Fig.~\ref{fig14} but for Ge NCs.}
\end{figure}

Considering 3~nm and 4~nm diameters, the results are shown in 
Figs.~\ref{fig14} and \ref{fig15} for Si and Ge NCs, respectively. 
Note that the 532~nm excitation results are 
qualitatively similar to those in intraband absorption, cf. 
Figs.~\ref{fig12} and \ref{fig13}. This is expected on the grounds of small excess 
energy for this case. Some general trends can be extracted from these results. 
First of all, the conduction band absorption is in general 
smooth over a wide energy range. On the other hand the valence band absorption 
contains pronounced absorption at several narrow energy windows mainly below 1~eV and 
they get much weaker than the conduction band absorption in the remaining 
energies. As the excitation energy increases the absorption coefficient per excited 
carrier in general decreases. 
In connection to silicon photonics, we should point out that the excited-state absorption 
is substantial including the important 1.55~$\mu$m fiber optics communication wavelength.
These results provide a more comprehensive picture than the reported experimental
measurements~\cite{elliman,forcales,trojanek} which are usually obtained at a single 
energy of the probe beam.
Finally, it needs to be mentioned that for both intraband and excited-state absorptions 
displayed in Figs.~\ref{fig12} to \ref{fig15}, the high energy parts will be masked by 
the interband transition whenever it becomes energetically possible.

\renewcommand{\baselinestretch}{1.5}
\begin{table*}
%Table-Excited State Pump Energies
\caption{The excited-state energies of the carriers within the valence and conduction states 
under three different interband pump energies for \emph{nc}-Si and \emph{nc}-Ge. The energies 
are  given in eV and measured from the HOMO and LUMO, respectively.}
\begin{ruledtabular} 
\begin{tabular}{c c c c c c c}
	&   &  \emph{nc}-Si   &	 &  & \emph{nc}-Ge    &  \\
\cline{2-4} \cline{5-7}
Pump & $D=3$~nm &         &  $D=4$~nm	        &	$D=3$~nm  &      &  $D=4$~nm	      \\
\hline                                                      
  2.33	    &	0.197, 0.021	&	& 0.021, 0.551	&	0.103, 0.643	&	& 0.211, 0.663  \\
  3.50	    &	0.000, 1.432	&	& 0.316, 1.440    &	0.400, 1.508    &	& 1.141, 0.892  \\
  4.66	    &	0.188, 2.414	&	& 0.713, 2.218    &	0.511, 2.551    &	& 1.360, 1.853  \\
\end{tabular}
\end{ruledtabular} 
\end{table*}
\renewcommand{\baselinestretch}{1}


\section{Conclusions}
The subject of Si and Ge NCs has become an established research field. A fundamental 
process such the direct photon absorption deserves further investigation from a number 
of perspectives. In this theoretical study, we consider the interband, intraband and 
excited-state absorption in embedded Si and Ge NCs of various sizes. For this purpose, 
we developed an atomistic pseudopotential electronic structure tool, the results of which 
agree very well with the published data. It is further observed that the HOMO for both 
Si and Ge NCs and the LUMO for Si NCs display oscillations with respect to size 
among different representations of the $C_{3v}$ point group to which these spherical NCs
belong. Our detailed investigation of the interband absorption reveals the importance of 
surface polarization effects that significantly reduce the absorption when included. 
This reduction is found to increase with decreasing NC size or with increasing permittivity 
mismatch between the NC core and the host matrix. These findings should be taken into account for 
applications where the absorption is desired to be either enhanced or reduced. 
For both NC types the deviation from sphericity shows no pronounced effect 
on the interband absorption. Next, the intraband process is considered which has potential applications on 
mid- and near-infrared photodetection. The intravalence band absorption is stronger compared to 
intraconduction band especially below 1~eV. Finally, we demonstrate that optical pumping 
introduces a new degrees of freedom to intraband absorption. This is studied under the title 
of excited-state absorption. Our major finding is that the excited-state absorption 
is substantial including the important 1.55~$\mu$m fiber optics communication wavelength.
Within the context of achieving gain and lasing in these NCs, excited-state absorption 
is a parasitic process, however, it can acquire a positive role in a different application.

\begin{acknowledgments}
The author is grateful to Aykutlu D\^ana for his suggestion of the intersubband absorption 
and its photodetector applications.
This work has been supported by the European FP6 Project SEMINANO with the 
contract number NMP4 CT2004 505285 and by the Turkish Scientific and Technical Council 
T\"UB\.ITAK with the project number 106T048. The computational resources are supplied in 
part by T\"{U}B\.ITAK through TR-Grid e-Infrastructure Project.

\end{acknowledgments}

\appendix
\begin{table*}
%Table-EPM
\caption{Parameters of the pseudopotential form factors of Si, Ge and their wide band-gap 
matrices. $a_0$ is the lattice constant. See text for the units.}
\begin{ruledtabular} 
\begin{tabular}{l c c c c c c c}
 			& $a_0$ (\AA)& $a_1$ & $a_2$ & $a_3$ & $a_4$ & $a_5$ & $a_6$  \\
\hline                                                      
 Si   		& 5.43 & 1.5708 & 2.2278 & 0.606 & -1.972 & 5.0 & 0.3   \\
Matrix-Si  & 5.43 & 1.5708 & 2.5    & 0.135 & -13.2  & 6.0 & 0.3   \\
 Ge   		& 5.65 & 0.7158 & 2.3592 & 0.74  & -0.38  & 5.0 & 0.3   \\
Matrix-Ge  & 5.65 & 0.4101 & 2.7    & 0.07  & -2.2   & 5.0 & 0.3   \\
\end{tabular}
\end{ruledtabular} 
\end{table*}
\section*{Appendix: Some technical details on the LCBB implementation}
In this section we first provide the details on the pseudopotential form factors of the bulk 
Si, Ge and their associated wide band-gap matrices to substitute for SiO$_2$.
We use the local empirical pseudopotentials for Si and Ge developed by Friedel, 
Hybertsen and Schl\"uter.~\cite{friedel} They use the following functional form for the 
pseudopotential form factor at a general wave number $q$:
$$
%\label{Vpsp}
V_{\mbox{\tiny{PP}}}(q) = \frac{a_1\left( q^2-a_2\right)}{e^{a_3\left( q^2-a_4\right)}+1} 
\left[ \frac{1}{2}\tanh \left( \frac{a_5-q^2}{a_6}\right) +\frac{1}{2}\right]\, .
$$
Using the parameters supplied in Table~III, the pseudopotential form factors come out in Rydbergs 
and the wave number in the above equation should be taken in atomic units (1/Bohr radius).
Another important technical remark is about the EPM cut off energies.
We observed that even though the EPM band energies (i.e., eigenvalues) converge reasonably well with cut
off energies as low as 5-10 Ry, the corresponding Bloch functions (i.e., eigenvectors) require substantially
higher values to converge.~\cite{bulutay06} The results in this study are obtained using 14 and 16 Ry for Si, 
Ge, respectively.

Finally, some comments on the LCBB basis set construction is in order. We only employ the bulk bands of 
the core material. The bulk band indices are chosen from the four valence bands an the lowest three to 
four conduction bands; usually these are not used in conjunction but separately for the NC valence and 
conduction states, respectively. The basis set is formed from a sampling over a three-dimensional rectangular 
grid in the reciprocal space centered around the $\Gamma$-point. Its extend is determined by the full 
coverage of the significant band 
extrema, such as conduction band minima of Si at the six equivalent 0.85~X points. The final LCBB basis set 
typically contains some ten thousand members.


\begin{thebibliography}{99}
\bibitem{ossicini}S. Ossicini, L. Pavesi, and F. Priolo, \textit{Light Emitting Silicon for Microphotonics}, 
Springer-Verlag, Berlin, (2004).
\bibitem{pavesi00}L. Pavesi, L. Dal Negro, C. Mazzoleni, G. Franz\'o, and F. Priolo, Nature {\bf 408,} 440 (2000).
\bibitem{walters}R. J. Walters, G. I. Bourianoff, and H. A. Atwater, Nat. Mater. {\bf 4,} 143 (2005).
\bibitem{rong}H. Rong, A. Liu, R. Jones, O. Cohen, D. Hak, A. Fang, and M. Paniccia, 
Nature (London) {\bf 433,} 292 (2005); H. Rong, R. Jones, A. Liu, O. Cohen, D. Hak, A. Fang, and M. Paniccia, 
Nature (London) {\bf 433,} 725 (2005).
\bibitem{wolkin}M. V. Wolkin, J. Jorne, P. M. Fauchet, G. Allan, and C. Delerue, Phys. Rev. Lett.
 {\bf 82,} 197 (1999).
\bibitem{zhou}Z. Zhou, L. Brus, and R. Friesner, Nano Lett.
 {\bf 3,} 163 (2003).
\bibitem{luppi}M. Luppi and S. Ossicini, Phys. Rev. B {\bf 71,} 035340 (2005).
\bibitem{sagnes}I. Sagnes, H. Halimaoui, G. Vincent, and P. A. Badoz, Appl. Phys. Lett.
 {\bf 62,} 1155 (1993). 
\bibitem{pv}M. Green, \textit{Third Generation Photovoltaics}, Springer-Verlag, Berlin, (2006).
\bibitem{furukawa}S. Furukawa, and T. Miyasato, Phys. Rev. B {\bf 38,} 5726 (1988).
\bibitem{kanemitsu}Y. Kanemitsu, H. Uto, Y. Masimoto, and Y. Maeda, Appl. Phys. Lett. 
{\bf 61,} 2187 (1992).
\bibitem{takeoda}S. Takeoda, M. Fujii, S. Hayashi, and K. Yamamoto, Phys. Rev. B
{\bf 58,} 7921 (1998).
\bibitem{wilcoxon99}J. P. Wilcoxon, G. A. Samara, and P. N. Provencio, Phys. Rev. B
{\bf 60,} 2704 (1999).
\bibitem{kovalev}D. Kovalev, J. Diener, H. Heckler, G. Polisski, N. K\"unzner, 
and F. Koch, Phys. Rev. B {\bf 61,} 4485 (2000).
\bibitem{wilcoxon01}J. P. Wilcoxon, P. P. Provencio, and G. A. Samara, Phys. Rev. B
{\bf 64,} 035417 (2001).
\bibitem{vasiliev}I. Vasiliev, S. \"O\u{g}\"ut, and J. R. Chelikowsky, Phys. Rev. Lett. {\bf 86,} 1813 (2001).
\bibitem{weissker02a}H.-Ch. Weissker, J. Furthm\"uller and F. Bechstedt, Phys. Rev. B {\bf 65,} 155327 (2002).
\bibitem{weissker02b}H.-Ch. Weissker, J. Furthm\"uller and F. Bechstedt, Phys. Rev. B {\bf 65,} 155328 (2002).
\bibitem{weissker03}H.-Ch. Weissker, J. Furthm\"uller and F. Bechstedt, Phys. Rev. B {\bf 67,} 245304 (2003).
\bibitem{ramos}L. E. Ramos, H.-Ch. Weissker, J. Furthm\"uller and F. Bechstedt, Phys. Status Solidi B 
{\bf 242,} 3053 (2005).
\bibitem{melnikov}D. Melnikov  and J. R. Chelikowsky, Solid State Commun. {\bf 127,} 361 (2003);
D. Melnikov  and J. R. Chelikowsky, Phys. Rev. B {\bf 69,} 113305 (2004).
\bibitem{tsolakidis}A. Tsolakidis and R. M. Martin, Phys. Rev. B {\bf 71,} 125319 (2005).
\bibitem{trani05}F. Trani, G. Cantele, D. Ninno, and G. Iadonisi, Phys. Rev. B {\bf 72,} 075423 (2005).
\bibitem{lin}Y. -Y. Lin and J. Singh, J. Appl. Phys. {\bf 96,} 1059 (2004).
\bibitem{qdip}V. Ryzhii, I. Khmyrova, V. Mitin, M. Stroscio, and M. Willander, Appl. Phys. Lett. 
 {\bf 78,} 3523 (2001).
\bibitem{malko}A. V. Malko, A. A. Mikhailovsky, M. A. Petruska, 
J. A. Hollingsworth, and V. I. Klimov, J. Phys. Chem. B {\bf 108,} 5250 (2004).
\bibitem{elliman}R. G. Elliman, M. J. Lederer, N. Smith, and B. Luther-Davies, Nuc. 
Instrum. Methods Phys. Res., Sect. B {\bf 206,} 427 (2003).
\bibitem{trojanek}F. Troj\'anek, K. Neudert, M. Bittner, and P. Mal\'y, Phys. Rev. 
B {\bf 72,} 075365 (2005).
\bibitem{forcales}M. Forcales, N. J. Smith, and R. G. Elliman, J. Appl. Phys.
B {\bf 100,} 014902 (2006).
\bibitem{zunger}A. Zunger, Phys. Status Solidi A {\bf 190,} 467 (2002).
\bibitem{martin-book}R. M. Martin, \textit{Electronic Structure}, Cambridge University Press, 
Cambridge (2004).
\bibitem{ogut}S. \"O\u{g}\"ut, J. R. Chelikowsky, S. G. Louie, Phys. Rev. Lett. {\bf 79,} 1770 (1997).
\bibitem{delerue04}C. Delerue and M. Lannoo, \textit{Nanostructures: Theory and Modelling}, 
Springer-Verlag, Berlin, (2004).
\bibitem{ren}S. Y. Ren, Phys. Rev. B {\bf 55,} 4665 (1997); S. Y. Ren, Solid State Comm. 
{\bf 102,} 479 (1997).
\bibitem{niquet}Y. M. Niquet, G. Allan, C. Delerue, and M. Lannoo, Appl. Phys. Lett.
{\bf 77,} 1182 (2000).
\bibitem{wang94a}L. -W. Wang and A. Zunger, J. Chem. Phys. {\bf 100,} 2394 (1994).
\bibitem{wang97}L. -W. Wang, A. Franceschetti and A. Zunger, Phys. Rev. Lett. {\bf 78,} 2819 (1997).
\bibitem{wang99}L. -W. Wang and A. Zunger, Phys. Rev. B {\bf 59,} 15806 (1999).
\bibitem{ninno85}D. Ninno, K. B. Wong, M. A. Gell and M. Jaros, Phys. Rev. B {\bf 32,} 2700 (1985).
\bibitem{ninno86}D. Ninno, M. A. Gell and M. Jaros, J. Phys. C: Solid State Phys. {\bf 19,} 3845 (1986).
\bibitem{botti01}S. Botti and L. C. Andreani, Phys. Rev. B {\bf 63,} 235313 (2001).
\bibitem{botti04}S. Botti, N. Vast, L. Reining, V. Olevano and L. C. Andreani, Phys. 
Rev. B {\bf 70,} 045301 (2004)
\bibitem{chirico}F. Chirico, A. Di Carlo and P. Lugli, Phys. Rev. B {\bf 70,} 045314 (2001).
\bibitem{bulutay07}C. Bulutay, Physica E {\bf 38,} 112 (2007).
\bibitem{pecvd}F. Iacona, G. Franz\'o, C. Spinella, J. Appl. Phys. {\bf 87,} 1295 (2000).
\bibitem{note1}An intermediate value between 0 and 1 can be used for the alloys or modeling the 
interface region. However, in this work we set them to either 1 or 0.
\bibitem{friedel}P. Friedel, M. S. Hybertsen, and M. Schl\"uter, Phys. Rev. B {\bf 39,} 7974 (1989).
\bibitem{chelikowsky77}J. R. Chelikowsky, Solid State Commun. {\bf 22,} 351 (1977).
\bibitem{jackson}J. D. Jackson, \textit{Classical Electrodynamics}, 2nd edition, Wiley, New York (1975).
\bibitem{trani07}F. Trani, D. Ninno, and G. Iadonisi, Phys. Rev. B {\bf 75,} 033312 (2007).
\bibitem{mahan}G. D. Mahan, Phys. Rev. B {\bf 74,} 033407 (2006).
\bibitem{delley}B. Delley and E. F. Steigmeier, Phys. Rev. B {\bf 47,} 1397 (1993).
\bibitem{reboredo01}F. A. Reboredo and A. Zunger, Phys. Rev. B {\bf 63,} 235314 (2001).
\bibitem{onida}G. Onida, L. Reining, and A. Rubio, Rev. Mod. Phys. {\bf 74,} 601 (2002).
\bibitem{ehrenreich}H. Ehrenreich and M. H. Cohen, Phys. Rev. {\bf 115,} 786 (1959).
\bibitem{hybertsen}M. S. Hybertsen, Phys. Rev. Lett. {\bf 72,} 514 (1994).
\bibitem{wang94b}L. -W. Wang and A. Zunger, J. Phys. Chem. {\bf 98,} 2158 (1994).
\bibitem{reboredo00}F. A. Reboredo, A. Franceschetti, and A. Zunger, Phys. Rev. B {\bf 61,} 13073 (2000).
\bibitem{delerue00}C. Delerue, M. Lannoo, and G. Allan, Phys. Rev. Lett. {\bf 84,} 2457 (2000).
\bibitem{garoufalis}C. S. Garoufalis, A. D. Zdetsis, and S. Grimme, Phys. Rev. Lett. 
{\bf 87,} 276402 (2001).
\bibitem{desousa}J. S. de Sousa, J.-P. Leburton, V. N. Freire, and E. F. da Silva Jr., 
Appl. Phys. Lett. 87, 031913 (2005).
\bibitem{mimura}A. Mimura, M. Fujii, S. Hayashi, D. Kovalev, and F. Koch, Phys. Rev. B 
{\bf 62,} 12625 (2000).
\bibitem{bulutay06}C. Bulutay, Turk. J. Phys. {\bf 30,} 287 (2006).
\end{thebibliography}
\end{document}